\documentstyle{article}
\newfont{\gothic}{eufm10 scaled\magstep 1}
\def\ss{\hbox{\gothic s}}
\def\JJ{\hbox{\gothic J}}
\def\EE{\hbox{\gothic E}}
\def\FF{\hbox{\gothic F}}
\def\GG{\hbox{\gothic G}}
\begin{document}
\begin{flushright} ANL-HEP-PR-92-89 \\  Miami TH-2-92 \\
\hfil hepth@xxx/9210060           \end{flushright}

\vskip 1.5in

\noindent {\bf SUPERSYMMETRY AND THE NONLOCAL YANGIAN DEFORMATION SYMMETRY}
\vskip 0.5in
\noindent \hskip 1in Thomas Curtright\footnote{Work supported by the NSF grant
PHY-92-09978.}$^{\S}$ and Cosmas Zachos\footnote{ Work supported by
the U.S.~Department of Energy, Division of High Energy Physics, Contract
W-31-109-ENG-38. }$^{\P}$

\vskip 0.25in

\noindent \hskip 1in $^{\S}$ Department of Physics, University of Miami,
Box 248046,

\noindent \hskip 1in Coral Gables, Florida 33124, USA~~~~~~~~~~~~{\sf
(tcurtright@umiami)}

\noindent \hskip 1in $^{\P}$ High Energy Physics Division,
Argonne National Laboratory,

\noindent \hskip 1in Argonne, IL 60439-4815, USA
{}~~~~~~~~~~~~~~~~~{\sf (zachos@hep.anl.gov)}

\vskip 0.5in

\noindent {\bf Abstract:}~~In the quantized two-dimensional non-linear
supersymmetric $\sigma$-model, the supercurrent supermultiplet,
which contains the energy-momentum tensor, is transformed by the nonlocal
symmetry of the model into the isospin current supermultiplet.
This effect incorporates supersymmetry into the known
infinite-dimensional Yangian deformation symmetry of plain $\sigma$-models,
leads to precisely the same nontrivial extension of the two-dimensional
super-Poincar\'e group as found previously for the Poincar\'e group,
and thus determines the theory's mass spectrum.
A generalization to all higher-order nonlocal charges is conjectured such
that their generating function, the so-called ``master charge'',
has a definite Lorentz spin which depends on the spectral parameter.

\vskip 0.7in

{\sl \noindent We contribute this  to Prof.~Biedenharn's Festschrift in
recognition of his appreciation of symmetries in physics and his timely
interest  in promising features of Hopf algebras.
\begin{flushright} Allez en avant, et la foi vous viendra.~~(J.~D' Alembert)
\end{flushright} }

One of the more productive proposals for probing the nonperturbative
structure of field theory is the utilization of nonlocal disorder variables
\cite{Polyakov'80}, predicated on the celebrated  nonlocal symmetries
\cite{LuscherPohlmeyer} of solvable two-dimensional models, such as the
nonlinear
$\sigma$-model, the Gross-Neveu model, or their supersymmetric combination.
Such nonlocal symmetries have provided  determinations of the S-matrices of the
respective models  \cite{LuscherZamo,Witten}. More recently, in the wake of
the advent of ``Quantum Deformation" Hopf algebra applications in field theory,
Bernard \cite{Bernard'91} noted that the algebraic structure of these
nonlocal symmetries, in fact, comprises a ``Yangian" algebra
\cite{Drinfel'd'85,Drinfel'd'88}, i.e.~a non-co-commutative coproduct
deformation of affine Lie algebras.
By dint of its nontrivial coproduct action on composite
states, the symmetry evades the Coleman-Mandula theorem, and,
due to a remarkable but simple quantum effect,
provides a nontrivial extension of the Poincar\'e group
\cite{Bernard'91,Belavin'92,BernardLeClair'92}:  the lowest order conserved
nonlocal charge $Q_{(1)}$ is not invariant under a Lorentz boost.
This engenders additional ``kinematic" constraints on the physical states
of the underlying theory which Belavin \cite{Belavin'92}
subsequently applied to cogently rederive the mass
spectrum of the SU(N) nonlinear $\sigma$-model.

Actually, such a quantal extension of the Poincar\'e group was to be
expected on the basis of known results for local lagrangean field theories.
On the one hand, as must be the case in a local field theory, any such
extension of the Poincar\'e charges, as that in question,
would be underlied by a current-multiplet structure involving nonlocal
transformations of the local energy-momentum tensor.
Therefore any known nonlocal transformation of
the {\sl trace} of the energy-momentum tensor
involving at most a single derivative of a local field
would suggest that the untraced energy-momentum tensor was similarly
transformed under the action of the nonlocal charge
and thereby provide a signal for an extension of the Poincar\'e algebra.
On the other hand, the trace of the energy-momentum tensor encodes
the quantum violation of dilation invariance through the trace anomaly
\cite{TraceAnom} which is proportional to the lagrangean density (with the
renormalization group $\beta$-function appearing as the coefficient of
proportionality),   and it has been known for quite a
while (e.g. in the context of the Gross-Neveu model \cite{CurtrightZachos'81})
that the first nonlocal charge transforms the lagrangean density into the
divergence of the local axial ``isospin" internal symmetry current.
It follows that the nonlocal transform of the energy-momentum tensor itself
involves the isospin current, and a unique, local, conserved result consisting
of skew derivatives of the isospin current is immediately found
which reproduces the correct trace. Indeed, just such a result for the
transformation of $T_{\mu\nu}$ was derived  from known form-factors of the
energy-momentum tensor and the isospin current for a particular
model  \cite{LeClairSmirnov'92}.
This  result then gives the aforementioned extension of the Poincar\'e group.

There is a point to be clarified, however.  The previous remarks would suggest
that the energy-momentum tensor transform under the action of the nonlocal
charge into the renormalization group $\beta$-function multiplying skew
derivatives of the isospin current, and hence that the $\beta$-function
appear in the  Poincar\'e extension.  Evidently \cite{Bernard'91} this is not
the case.   The coefficient which appears in the extension is
an exact ${\cal O}(\hbar)$ quantity resulting from ``topological"
considerations. We discuss this more later.

Results for lagrangean field theory also suggest a conjecture for the
all-orders (in the spectral parameter) nonlocal effect on the energy-momentum
tensor. Specifically, consistently to inputs such as the all-orders
transformation of ${\cal L}$  previously
obtained for the Gross-Neveu model \cite{CurtrightZachos'81}, we propose that
the complete conserved ``master-charge" (the generating
functional of {\sl all} nonlocal charges) acting on the energy-momentum tensor
yields skew derivatives of the ``master current".
Integrated, this conjectured relation implies that the master charge possesses
definite Lorentz spin.

The natural question arises whether, in supersymmetric models, this
quantal algebra generalizes to involve supersymmetry as well,
and whether the above-mentioned extension of the Poincar\'e group can be
stretched further to include nontrivial action among yet more conserved
(graded) generators. We answer these two questions in the affirmative, and the
negative, respectively.

Specifically, we find that, for the supersymmetric nonlinear $\sigma$-model,
the lowest nonlocal charge transforms the entire supercurrent
supermultiplet into the vector ``isospin" current supermultiplet, just as it
does to components thereof (viz.~the energy-momentum tensor into skew
derivatives of the isospin current) in the plain $\sigma$-model or Gross-Neveu
cases. This follows from requirements of consistent algebraic closure of
supersymmetry, provided that the supercharge commutes with the nonlocal charge
in the quantum theory as it does in the classical limit\footnote{
This type of algebraic closure evokes the Wess-Zumino anomaly consistency
conditions, i.e.~Lie-algebraic closure on the transform of the lagrangean.}.

However, in contrast to the extension of the algebra of conserved charges
(boosts) in the plain Poincar\'e case, there are no additional conserved
(graded) charges acting nontrivially on the nonlocal charge and therefore no
additional ``kinematic" constraints in the super-Poincar\'e case, unless the
quantum theory is also conformally invariant.  As a consequence, the
supersymmetric $\sigma$-model  is constrained by the very same Yangian
relations as each of its non-supersymmetric components, with only the usual
Fermi-Bose degeneracy  resulting from the supersymmetry \cite{Witten}.

We now expand on all these points more explicitly for the supersymmetric
$\sigma$-model \cite{Witten}.
The O(N)-invariant supersymmetric $\sigma$-model involves $N$ real scalar
fields $n^a$ and $N$ Majorana spinors $\psi^a$, constrained by
\begin{equation}
n^a n^a =1~, ~~~~~~~~~~~~~~~~~ n^a \psi^a=0 ~,
\end{equation}
and described by the lagrangean
\begin{equation}
{\cal L} ={\textstyle\frac12} ~\partial_\mu n^a \partial ^\mu n^a+
{\textstyle \frac 12}~i\,\overline{\psi} ^a\gamma ^\mu \partial _\mu \psi^a
+{\textstyle \frac 18}~g\left( \overline{\psi} ^a\psi ^a\right) ^2 ~.
\end{equation}
Beyond O(N) invariance, the model is further invariant under supersymmetry
transformations
\begin{equation}
\delta n^a=\bar\epsilon \psi^a~, ~~~~~~~~~~~~~~~~ \delta\psi^a=
[-i~/\kern-.5em\partial n^a + {\textstyle \frac 12}~ n^a
\left( \overline{\psi} ^b\psi ^b\right) ]~\epsilon~,
\end{equation}
for which the supercurrent is
\begin{equation}
S_\mu=-i (\partial_\nu n^a)~ \gamma^\nu \gamma_\mu \psi^a ~.
\end{equation}
The conserved Noether currents $J_\mu^{ab}$ generating the O(N) rotations are
split into fermion ($2~B_\mu^{ab}$) and boson ($A_\mu^{ab}$)  contributions:
\begin{equation}
\label{isospin} J_\mu^{ab}= A_\mu^{ab}+2~B_\mu^{ab}~, ~~~~~~~~~~~~~~~~~~
A_\mu^{ab}=  2\; n^a \stackrel{\leftrightarrow} {\partial}_\mu n^b~,
{}~~~~~~~~~~~ B_\mu^{ab}=  -i ~\overline{\psi}^a \gamma_\mu \psi^b~.
\end{equation}
Using the equations of motion, it is straightforward to see that
\cite{CurtrightZachos'80}
\begin{equation}
\label{eee} {\partial}^\mu A_\mu^{ab}=  A^{\mu~ ac} B_\mu^{cb}-B^{\mu~ ac}
A_\mu^{cb}~,
\end{equation}
\begin{equation}
{\partial}^\mu B_\mu^{ab}=-{\textstyle \frac 12}~A^{\mu~ac} B_\mu^{cb}
+{\textstyle \frac 12}~B^{\mu~ ac}  A_\mu^{cb}~,
\end{equation}
\begin{equation}
\epsilon^{\mu\nu}{\partial}_\mu A_\nu^{ab}=
-\epsilon^{\mu\nu} A_\mu^{ac} A_\nu^{cb}~,
\end{equation}
\begin{equation}
\label{eef} \epsilon^{\mu\nu}{\partial}_\mu B_\nu^{ab}=
-\epsilon^{\mu\nu} B_\mu^{ac} B_\nu^{cb}-{\textstyle \frac
12}~\epsilon^{\mu\nu}
( A_\mu^{ac} B_\nu^{cb}+ B_\mu^{ac} A_\nu^{cb})~.
\end{equation}
One may use these to prove conservation of $J^{ab}_{(0)~\mu}\equiv J_\mu^{ab}$
and the properties of $K_\mu$  and $C_\mu$ introduced below.

To establish all this, recall the
identities
\begin{equation}
\gamma^\mu  \gamma^\nu= g^{\mu\nu}  \hbox{{1}\kern-.25em\hbox{l}}
+ \epsilon^{\mu\nu} ~\gamma_P~,
{}~~~~~~~~~~~~~~~~~~~~~~~~~~\gamma^\mu =\gamma_P~ \epsilon^{\mu\nu}
\gamma_\nu~,
\end{equation}
\begin{equation}
\epsilon^{\kappa\lambda} \epsilon^{\mu\nu}= g^{\kappa\nu}g^{\lambda\mu}-
g^{\kappa\mu}g^{\lambda\nu}~,
{}~~~~~~~~~~~~~~~~~~~~
g^{\kappa\lambda }\epsilon^{\mu\nu}+ g^{\kappa\mu}\epsilon^{\nu\lambda}+
g^{\kappa\nu}\epsilon^{\lambda\mu}=0~,
\end{equation}
where the pseudoscalar matrix is defined as
\begin{equation}
\gamma_P\equiv \gamma^0 \gamma^1 ~,~~~~~~~~~~~~~~~~~~~
\gamma_P^{~2}= \hbox{{1}\kern-.25em\hbox{l}} ~.
\end{equation}
Also recall the Majorana transposition properties
\begin{equation}
\overline{\psi} \phi = \overline{\phi}\psi~,~~~~~~~~~~~~~~~~~~~~~~~~
\overline{\psi} \gamma_\mu  \phi =-\overline{\phi}\gamma_\mu \psi~
,~~~~~~~~~~~~~~~~~~
\overline{\psi} \gamma_P \phi =-\overline{\phi} \gamma_P\psi~,
\end{equation}
as well as the Fierz rule
\begin{equation}
\varphi~ \overline{\psi}\phi =
-{\textstyle \frac12}(\phi~\overline{\psi}\varphi+
\gamma_\mu \phi~ \overline{\psi} \gamma^\mu \varphi+
\gamma_P\phi~\overline{\psi} \gamma_P\varphi)~.
\end{equation}

The full set of nonlocal conservation laws is neatly described using the
methods in \cite{CurtrightZachos'80} (see also \cite{Polyakov'80,Zachos'80}).
Introduce a dual boost spectral parameter $\theta$ to define
\begin{equation}
\label{C}C_\mu ={\textstyle \frac 12}\left( 1-\cosh \theta \right) A_\mu -
{\textstyle \frac 12}\sinh\theta \;\tilde A_\mu +{\textstyle \frac 12}\left(
1-\cosh 2\theta \right) B_\mu -{\textstyle \frac 12}\sinh 2\theta \;
\widetilde B_\mu \;,
\end{equation}
\begin{equation}
\label{K}K_\mu =\cosh \theta \;A_\mu +\sinh \theta \;\tilde A_\mu +2\cosh
2\theta \;B_\mu +2\sinh 2\theta \;\widetilde B_\mu \;,
\end{equation}
where $\tilde A_\mu \equiv \epsilon _\mu ^{\;\;\nu }\,A_\nu $ . (To compare to
the spectral parameter $\kappa$ in
\cite{CurtrightZachos'80}, use $\cosh \theta =\frac{1+\kappa ^2}{1-\kappa ^2}$
 and $\sinh \theta =
\frac{2\kappa }{1-\kappa ^2}$.)  Now, as is the case for the classical
supersymmetric $\sigma $-model on-shell (readily checked by virtue of
Eqs(\ref{eee}-\ref{eef})),
\begin{equation}
\label{ZeroCurvC}\;\left( \partial ^\mu +C^\mu \right) \widetilde C_\mu =0\;,
\end{equation}
\begin{equation}
\label{ZeroCurvK}\partial ^\mu K_\mu +[C^\mu ,K_\mu ]=0\;.
\end{equation}
These allow construction of  a conserved nonlocal ``master current''
\begin{equation}
\label{NMC}\JJ^\mu (x)=\chi ^{-1}(x)\;K^\mu (x)\;\chi (x)\equiv
 \sum_{n=0}^\infty \theta^n ~J_{(n)} ^{\mu}\;,
\end{equation}
where $\chi $ is the path-ordered exponential solution to
\begin{equation}
\label{PathOrdExp}\partial _\mu \,\chi ^{ab}(x)=-C_\mu ^{ac}\,\chi
^{cb}(x)\;.
\end{equation}
This solution is possible by virtue of the consistency condition
Eqn(\ref{ZeroCurvC}), and amounts to Polyakov's disorder variable
\cite{Polyakov'80,CurtrightZachos'80}
\begin{equation}
\label{realPathOrdExp} \chi (x)=P\exp\Bigl(-\int^x_{-\infty}dy~ C_1(y,t) \Bigr)
\;.
\end{equation}

The master current acts as the generating functional of all
 currents $J_{(n)} ^{\mu}$ (separately) conserved order-by-order in $\theta$;
e.g.~the lowest two orders yield the local and first nonlocal currents,
respectively:
$$
\JJ_\mu= \Bigl( A_\mu(x)+2B_\mu(x)\Bigr) ~+
$$
\begin{equation}
+~\theta~ \Biggl( \tilde A_\mu(x)+4\widetilde B_\mu(x)~+
{\textstyle \frac 12} \Bigl[A_\mu(x)+2B_\mu(x)~,~\int_{-\infty}^x \kern-1.2em
 dy (A_0(y)+2B_0(y))\Bigr] \Biggr)~+~{\cal O}(\theta^2)~.
\end{equation}
Integrating the nonlocal master current yields the conserved ``master charge''
\begin{equation}
\label{WeDon'tTakeAmericanExpress}\GG=\int_{-\infty }^{+\infty }dx\;\JJ
_0(x)\equiv \sum_{n=0}^\infty \theta^n ~Q_{(n)} \;.
\end{equation}
$Q_{(0)} $ is the conventional isospin charge, while $Q_{(1)} $ is the
well-known  lowest nonlocal charge.

As in \cite{LuscherZamo}, all of the above structure can be argued
to survive quantization (i.e.~we conjecture there are supersymmetric and
all-orders-in-$\theta$ versions of ``L\"uscher's Theorem"\cite{Bernard'91}) and
to admit a non-perturbative formulation.  However, further work on this is
necessary before a rigorous proof is available.

We now proceed to give the action of the lowest nonlocal charge $Q_{(1)} $
on the supercurrent supermultiplet of the model. Recall the familiar
supercurrent multiplet \cite{West}, $(R,S_\mu ,T_{\mu \nu })$, where
$S_\mu$ is the supercurrent and $T_{\mu \nu }$ is the stress-energy tensor.
The supersymmetry transformations are given by
\begin{equation}
\label{delR}\delta R=\,-i\,\bar \epsilon \,\gamma ^\lambda \,S_\lambda \;.
\end{equation}
\begin{equation}
\label{delS}\delta S_\mu =-i\,\left( 2\,T_{\mu \lambda }\,\gamma ^\lambda
\,+\,i\,\epsilon _{\mu \lambda }\,\partial ^\lambda R\,\gamma _{P\,}\right)
\epsilon \,,
\end{equation}
\begin{equation}
\label{delT}\delta T_{\mu \nu }=-\, {\textstyle \frac 14}\,\bar \epsilon \,
\gamma _P\left(
\epsilon _{\mu \lambda }\,\partial ^\lambda \,S_\nu +\epsilon _{\nu \lambda
}\,\partial ^\lambda \,S_\mu \right) =-\,{\textstyle \frac 12}\,\bar
\epsilon \,\gamma
_P\epsilon _{\mu \lambda }\,\partial ^\lambda \,S_\nu ~.
\end{equation}
In Eqn(\ref{delT}),  $\partial _\lambda S^\lambda =0$ has been used in the
last equality.
Next, the ``vector" isospin current supermultiplet \cite{CurtrightZachos'80}
transforms under supersymmetry as
\begin{equation}
\label{delPhi}\delta \Phi ^{ab}=\,-i\,\partial _\lambda \left( \bar \epsilon
\,\gamma ^\lambda \,E^{ab}\right) \;,
\end{equation}
\begin{equation}
\label{delE}\delta E^{ab}=\left( i\,\gamma _P\gamma ^\lambda \,J_\lambda
^{ab}+\,\Phi ^{ab}\right) \epsilon \;,
\end{equation}
\begin{equation}
\label{delJ}\delta J_\mu ^{ab}=\epsilon _{\mu \lambda }\,\partial ^\lambda
\left( \bar \epsilon \,\,E^{ab}\right) \;.
\end{equation}
Classically, for the specific model under consideration, these components
are just : $E^{ab}=-2 \gamma_P~(n^a\psi^b- n^b\psi^a) $,
$~~\Phi^{ab}=2~\overline{\psi}^a\gamma_P \psi^b  $, and  $J_{\mu}^{ab}$ as in
Eqn(\ref{isospin}).
Finally, the first nonlocal transformation of the supercurrent multiplet
involves the     isospin current supermultiplet:
\begin{equation}
\label{Q1U}i\,[Q^{ab}_{(1)} ,R]=2~f\,\Phi ^{ab}\;,
\end{equation}
\begin{equation}
\label{Q1S}i\,[Q_{(1)}^{ab},S_\mu ]=2~f~\gamma _P\,\epsilon _{\mu \lambda
}\partial ^\lambda E^{ab}\;,
\end{equation}
\begin{equation}
\label{Q1T}i\,[Q_{(1)}^{ab},T_{\mu \nu }]=-\,\frac 12 f\,\left( \epsilon _{\mu
\lambda }\partial ^\lambda J_\nu ^{ab}+\epsilon _{\nu \lambda }\partial
^\lambda J_\mu ^{ab}\right) =-f\,\epsilon _{\mu \lambda }\partial ^\lambda
J_\nu ^{ab}\;.
\end{equation}
In Eqn(\ref{Q1T}), $\partial ^\lambda J_\lambda ^{ab}=0$ has been used in the
last equality. Also note that Eqn(\ref{Q1T}) agrees with Bernard's relation
\cite{Bernard'91} (and coincides with the more general unintegrated version of
ref.\cite{LeClairSmirnov'92}). The mass spectrum analysis of Belavin
\cite{Belavin'92} then carries through {\em mutatis mutandis} for the
supersymmetric model.

The above three commutators supertransform to a combination of themselves.
That is, supersymmetry dictates the form of essentially two out of three
relations on the basis of one, given the previous two supermultiplet
structures. Further observe that integrating Eqn(\ref{Q1S}) is consistent with
the assumption that the supercharge commutes with $Q_{(1)}$, given
``good" boundary conditions.

Now, what is $f$? It is not specified by supersymmetric consistency in these
abstract commutators.  Classically, however, $f=0$ is required. Since
$\gamma\cdot$Eqn(\ref{Q1S}) and the trace of Eqn(\ref{Q1T}) reduce to
\begin{equation}
\label{tQ1S}i\,[Q_{(1)}^{ab},\gamma \cdot S]=2~f~/\kern-.5em\partial E^{ab}\;,
\end{equation}
\begin{equation}
\label{tQ1T}i\,[Q_{(1)}^{ab},T_{\mu}^{ \mu }]
=-f\,\epsilon ^{\mu \lambda }\partial_\lambda J_\mu ^{ab}\;,
\end{equation}
and the trace of the energy-momentum tensor on the left-hand-side
of Eqn(\ref{tQ1T}) vanishes classically, but the divergence of the axial
current on the right-hand-side does not, it follows that the coefficient $f$
must vanish in the classical limit. Is this triviality maintained upon
quantization?

Quantum mechanically, the trace anomaly on the left-hand-side of
Eqn(\ref{tQ1T}) is nonvanishing, and, in fact, proportional to the
$\beta$-function times the lagrangean density.
(This general result, \cite{TraceAnom} and references therein, sometimes termed
``Minkowski's conjecture", has been substantiated in a broad array of
models.) Consequently, one might expect $f$ on the right-hand-side to be
proportional to the $\beta$-function (or at least be a function with the
same zeros as the $\beta$-function).
However, Bernard's derivation of $f=i\hbar$ is essentially ``topological"
and therefore should give an exact result for $f$, provided only that the
nonlocal current is well-defined by simply point-splitting the local currents.
At most, Bernard's result therefore may be identified with the one-loop
$\beta$-function in the plain $\sigma$-model.

Of course, there is an obvious way to reconcile all this.
Namely, the transformation of the lagrangean density may acquire
quantum corrections which compensate for the higher order terms in $\beta$.
This is a logical possibility which will have to be investigated elsewhere.
Here, instead, we wish to follow-up on the $\beta$-function suggestion and
consider theories with nontrivial renormalization group fixed-points.
As an example, consider the WZW model in the geometrostatic limit
\cite{CurtrightZachos'84}. In this limit, which is an infra-red fixed-point,
both the energy-momentum tensor trace vanishes and the axial isospin
current is conserved. Thus, in this model, it is {\sl not} necessary for the
nonlocal transformation of the trace to involve the $\beta$-function.
Nonetheless, we have considered the explicit form of the nonlocal current for
the WZW model  \cite{deVega},  and followed Bernard's topological argument. We
find the nonlocal transformation of the {\sl untraced} energy-momentum tensor
to coincide with Eqn(\ref{Q1T}), i.e. the one-loop $\beta$-function does not
appear.     Hence the nonlocal charge transforms the
energy-momentum tensor even in the geometrostatic limit.
The reader should recall that this limit is equivalent to free massless
fermions, for which one can readily construct a self-dual (light-like)
conserved isocurrent, $j_{\mu}^{ab}=-i ~\overline{\Psi}^a \gamma_\mu
(1\pm  \gamma_P)\Psi^b~$, and hence the conserved nonlocal current,
$j_{(1)}^{\mu}(x) \equiv j^{\mu}(x)\int^{x}_{-\infty}dy  j_0(y)$. There is no
need for adding to  $j_{(1)}^{\mu}$ the usual local contribution in this free
fermion construction.

We proceed to conjecture that the master charge obeys a simple commutation
relation with the energy momentum tensor:
\begin{equation}
\label{GT} i\,\left[\GG,T_{\mu \nu }\right] ={\textstyle\frac12}~\ss(\theta)\,
\left( \epsilon _{\mu
\lambda }\partial ^\lambda \JJ_\nu +\epsilon _{\nu \lambda }\partial
^\lambda \JJ_\mu \right) =\,\ss(\theta  )\,\epsilon _{\mu \lambda
}\partial ^\lambda \JJ_\nu \;.
\end{equation}
As before, supersymmetric consistency dictates
\begin{equation}
\label{GU}i\,[\GG,R]=-2\ss(\theta )\,\FF~,
\end{equation}
\begin{equation}
\label{GS}i\,[\GG,S_\mu ]=-2\ss(\theta )~ \gamma _P\,\epsilon _{\mu \lambda
}\partial ^\lambda \EE\;,
\end{equation}
where $\FF$ and $\EE$ are the other two components
of the master-current supermultiplet
generalizing Eqns(\ref{Q1U},\ref{Q1S}) to all orders in $\theta$; and
$\ss(\theta)\rightarrow  \theta f= i \theta $ as $ \theta \rightarrow 0$.
No constraints on the form of $\ss(\theta )$ are presently available beyond
this limit.

The Lorentz boost is given by
\begin{equation}
L=-\int_{-\infty }^{+\infty }dx\; x~T_{00} \qquad \qquad \qquad (t=0)~,
\end{equation}
so Eqn(\ref{GT}) yields
\begin{equation}
\label{Lorspin} [L,\GG]= i\ss(\theta )~ \GG~,
\end{equation}
assuming there is no trouble from the surface term upon integration by parts
on the r.h.s. (This appears plausible as $K_{\mu}$ and whence the master
current vanish at infinity.) Thus not only is $\GG$ not Lorentz invariant
as was the case for $Q_{(1)}$, but, as a consequence of Eqn(\ref{Lorspin}),
the master charge is seen to have Lorentz spin $i\ss(\theta)$,
just as light-cone translations have Lorentz-spin $\pm 1$.

It is important to note that there is no ``moment" of the supercurrent
which is conserved and which could thus enlarge or complicate Bernard's
extended Poincar\'e algebra, unless the quantum theory is also conformally
invariant. In that case the algebra enlarges to include conserved conformal
currents  $M_{\mu}=x\cdot\gamma~ S_{\mu}$, $D_{\mu}=x^{\nu} T_{\nu\mu}$, and
$N_{\nu\mu}=2x_{\nu} x^{\lambda}T _{\lambda\mu} - x^2 T_{\nu\mu} $,
as well as the conventional  2-d infinite conformal extension.

As indicated in the introduction, we finally provide some of the motivation for
our all-orders conjecture, which is our older work \cite{CurtrightZachos'81}
on the Gross-Neveu model---unfortunately not printed on acid-free paper.
Suppressing the bosons in the lagrangean and Eqs(\ref{C}-\ref{realPathOrdExp}),
yields the Gross-Neveu lagrangean
\begin{equation}
\label{L}{\cal L}={\textstyle\frac12}
\,i\,\overline{\psi}^a\gamma ^\mu \partial _\mu \psi
^a-{\textstyle\frac18} \,g\,B_\mu ^{ab\,}B^{\mu \,ab}=
{\textstyle\frac12} \,i\,\overline{\psi}^a\gamma
^\mu \partial _\mu \psi ^a+{\textstyle\frac18}
\,g\left( \overline{\psi}^a\psi ^a\right) ^2~.
\end{equation}
It can be easily demonstrated that, on shell,
\begin{equation}
\epsilon _{\mu \lambda }\partial ^\mu \chi ^{-1}\;
\partial ^\lambda \chi =
{1-\cosh 2\theta   \over 2}~
\chi ^{-1}\;B^\mu\cdot \tilde B^\mu \;\chi =
{1-\cosh 2\theta   \over 4} ~\epsilon _{\mu \lambda }\partial ^\lambda
\JJ^\mu (x)~.
\end{equation}
Using ordered-exponential solutions $\chi[x,-\infty )$
of a somewhat different ordering prescription and a corresponding (but not
identical) master-charge \cite{CurtrightZachos'81}, this lagrangean was
classically shown to transform under the full action of the master  charge
according to
\begin{eqnarray}
\label{GL}i\,\left[ \kappa \widehat{\GG^{ab}},~{\cal L}\right]
&=& g^2\,\epsilon ^{\mu \nu }\,\partial _\mu \,\chi ^{ac}(\infty
,x]\,\partial _\nu \,\chi ^{cb}[x,-\infty )  \nonumber \\
 &=& g^2\,\epsilon ^{\mu \nu }\,\partial _\mu \,\chi ^{ca}[x,-\infty
)\,\partial _\nu \,\chi ^{cb}[x,-\infty )\;.
\end{eqnarray}
Up to an over-all coefficient (with quantum corrections?), this is readily
identified with the trace of our conjectured Eqn(\ref{GT}).

\end{document}